\documentclass[11pt,leqno,twoside]{article}
\usepackage{srcltx}
\usepackage{amsmath}
\usepackage{amsfonts}
\usepackage{amssymb}
\usepackage{amsthm}
\usepackage{latexsym}
\usepackage{afterpage}
\usepackage{cite}
\usepackage[dvips]{graphicx}
\usepackage{graphics}
\usepackage{array}
\usepackage{psfrag}
\author{Tassilo Ott\\
\small ott@physik.hu-berlin.de\\
\small Institut f\"ur Physik der Humboldt-Universit\"at zu Berlin\\
\small Invalidenstra\ss e 110\\
\small D-10115 Berlin, Germany}
\date{}
\title{\vbox{\hbox{\vspace{2cm}}{Gauge Invariant Perturbation Analysis for Quintessence with an Exponential Potential}}}
\begin{document}
\vspace{4cm}
\maketitle
\vskip -9.0cm
 \hspace{12.5cm} \hbox{\vbox{HU-EP-01/07~\\astro-ph/0102448}}
\vskip 9.0cm
\begin{abstract}
\vspace{0.5cm}
\noindent Gauge invariant adiabatic perturbations in the
cosmology of a scalar field $\varphi(\vec x,t)$ gravitationally
coupled to a perfect fluid (radiation or matter) are investigated.
The potential of the scalar field is of exponential form. This
model recently has received much attention in the context of
`quintessence' due to new observations of high redshift
supernovae indicating an accelerating universe. The perturbations
are treated in a gauge invariant formalism to systematically
avoid the residual gauge modes in synchronous gauge. Two
important attractor solutions of the unperturbed equations are
investigated: the first corresponds to the case where the energy
density of the unperturbed field $\rho_{(\varphi)}$ dominates the
cosmology after a short time, the second to a solution with a
time independent ratio $\rho_{(\varphi)}/\rho_{\rm hyd}$. The
main question addressed in this paper is if the inhomogenous
fluctuations can be ignored as often argued or not. Indeed, in
some important cases the perturbations do not die out with time
meaning that the ratio of energy density of the inhomogenous
fluctuations to the unperturbed part is constant after a while.
Consequences for microwave background radiation and structure
formation in the matter dominated era are considered in a
straightforward manner and it is concluded that for similar initial values
the anisotropy of the background radiation should be stronger in
this model than just for radiation. Furthermore, structure
formation should have happened slower than in standard cosmology.
\end{abstract}
\newpage
\section{Introduction}
In this paper gauge invariant perturbations in the context of
quintessence are investigated. Since Perlmutter et al. and Riess
et al. \cite{Riess:1998jm,Perlmutter:1998zf} have published their
results that the universe is accelerating instead of
decelerating, much work has been done to find the reason or at
least a correct kinematic description for this behaviour. One of
the best candidates is the so-called quintessence
matter-component that effectively can be described by purely
gravitational coupling of a scalar field to the standard matter
(radiation or non-relativistic matter)\cite{Caldwell:1998ii}.
There are many different suggestions for the form of the
quintessence-potential (see for instance
\cite{Caldwell:1998ii,Zlatev:1998tr,Sahni:1999qe,Peebles:1998qn}).
One of the first ones is the exponential
potential as it was originally discussed by Wetterich
\cite{Wetterich:1994bg, Wetterich:1988fm, Wetterich:1985wd} and
Ratra and Peebles \cite{Ratra:1988rm}. This potential will be
investigated in the form:
\begin{equation}
V(\varphi)=\hat{V}e^{-{\frac{a}{M}\varphi}} \ .
\end{equation}
In this definition M is the reduced planck mass ($M^2={M_{\rm
P}^2}/{16\pi}$) and $a$ the only free model parameter. The
amplitude $\hat{V}$ can be adjusted by a shift of the field
$\varphi$ and therefore does not play an important role. This
potential is not compatible with observation
\cite{Barreiro:1999zs} which could be changed in late cosmology,
but it still serves as a suitable toy model. The naturalness of
such models has been discussed by Hebecker et
al.\cite{Hebecker:2000zb}. Adiabatic Perturbations of this model
in synchronous gauge were already discussed by Ferreira and Joyce
\cite{Ferreira:1997au,Ferreira:1998hj} and
Amendola\cite{Amendola:1999dr,Amendola:1999er} but the influence
of inhomogenous perturbations onto the CMBR has been ignored.
\section{The ``background'' solutions}
From the Einstein equations, Energy-momentum conservation and the
Klein-Gordon equation the three field equations for just time
dependent quantities can be obtained for Robertson-Walker
background\cite{Wetterich:1994bg}:
\begin{align}
&{\ddot{\varphi}(t)+3H(t)\dot{\varphi}(t)+{{\partial
V(\varphi )} \over {\partial \varphi }}}=0\label{eq:fgu1}\ ,\\
&{1 \over 6{M}^2}\left(V(\varphi)+{\frac{1}{2}}\dot{\varphi}^2(t)+\rho(t)\right)=H^2(t)\label{eq:fgu2}\ ,\\
&\dot{\rho}(t)+3H(t)\Big(\rho(t)+p(t)\Big)=0\label{eq:fgu3} \ .
\end{align}
This work is in an inflationary context, so the density parameter
$\Omega\equiv{\rho_\mathrm{tot}}/{\rho_{\rm C}}$ is set to 1; an
extension to the other cases would be straightforward. An equation
of state is needed for the perfect fluid component and therefore
the convention
\begin{equation}
3\left(\rho +p\right)=n \rho \ .
\end{equation}
will be used such that $n=4$ gives a radiation dominated and $n=3$
a matter dominated universe. The nonlinear equations admit two
well-known ``attractor solutions''.

On the one hand this is the so-called ``tracker solution''(which
will be referred to as attractor solution I):
\begin{align}\label{eq:loes_attI}
& \bar{\varphi}(t)=\bar{\varphi}_0+{2M \over a}\ln{t \over
t_0} \ , & & H(t)={2 \over {nt}} \ , \\
& \bar\rho(t)=\frac{6a^2-3n}{a^2}M^2 H^2(t) \ ,   & &
V(t)=\frac{6n-n^2}{2a^2}M^2H^2(t) \ . \nonumber
\end{align}
The name arises because the time dependence of the scalar field
energy density tracks the behaviour of the perfect fluid. This
solution exists (meaning it is reached after a short period for
virtually all initial conditions) for $a^2 > n/2$ and there is no
induced cosmological constant and consequently it does not allow
for an accelerated expansion of the universe.

On the other hand there
is the solution (which will be called attractor solution II):
\begin{align} \label{eq:loes_attII}
\nonumber & \bar{\varphi}(t)=\bar{\varphi}_0+{2M \over a}\ln{t \over t_0} \ ,  & & H(t)={1 \over {a^2t}} \ , \\
&\bar\rho(t)=\rho_0\left(\frac{t}{t_0}\right)^{-\frac{n}{a^2}} \ , & &V(t)={2M^2 \over a^2}\left({3 \over a^2}-1\right){1
\over t^2} \ . \nonumber
\end{align}
It corresponds to the case where $\bar \rho >> V$ and exists for
$a^2<3$, but in the regime $3/2\leq a< \sqrt{3}$ for $n=3$ the
attractor solution I is dominant. The energy density part of a
dynamical cosmological constant in this case is non-vanishing:
\begin{equation}
\bar \rho_\Lambda=\frac{2M^2}{a^2t^2}\left(\frac{3}{a^2}-\frac{6}{n}\right) \ .
\end{equation}
Nucleosynthesis gives a constraint for $a$ during that period of $a > \sqrt{22}\approx 4.69$ \cite{Wetterich:1994bg} or
 $a > 4.5-5.5$ \cite{Ferreira:1998hj}.
 Therefore, a scenario seems to be appealing where $a$ is not
 fixed but decreasing with cosmic time. Then during radiation domination attractor
 solution I was valid with $a\sim5$ at the time of
 nucleosynthesis. At the beginning of matter domination, attractor
 solution I still was valid, but recently the value of $a$
 decreased below $a < \sqrt{3/2}$ and the universe ``switched
 over'' towards attractor solution II. This could explain the
 acceleration of the universe today through the effective cosmological constant.
\section{The gauge invariant perturbation equations}
The problem of residual gauge modes is known to cosmologists for
a long time (see for instance \cite{Weinberg:1972}), but has
attracted general attention in the early eighties when
Brandenberger et al. \cite{Brandenberger:1983vj} showed that
indeed many of the generally agreed results on perturbation
growth were not correct due to residual gauge modes. These cannot
be eliminated in general within synchronous gauge. A
sophisticated solution to this problem already had been offered
by Bardeen \cite{Bardeen:1980kt} a few years earlier. He first
took the approach to ´eliminate´ the gauge modes by making a
transformation from the original variables to new ones that
explicitly do not change under residual general coordinate
transformations. This approach is taken in this paper, following
Mukhanov et al.\cite{Mukhanov:1992me}. The most general scalar
metric perturbations are parameterised by the four variables
$\phi$, $\psi$, $B$ and $E$ within the line element
\begin{equation}
ds^2=\left[\left(1+2\phi\right)dt ^2-R(t)2B_{|i}dx^i dt-\Big[\left(1-2\psi\right)
\delta_{ij}+R^2(t)2E_{|ij}\Big] dx^i dx^j\right] \ .
\end{equation}
Here the symbol $|$ denotes the covariant derivative on the
hypersurface of constant time, $R$ is the scale factor and $t$
denotes cosmological time. Gauge invariant linear combinations of
the so-called metric potentials $\phi$ and $\psi$ can be
constructed by
\begin{align}
\Phi&=\phi+\frac{\partial}{\partial t}\left[\left(B-R\dot{E}\right)R\right]\ , \\
\Psi&=\psi-\dot R \left(B-R\dot E\right) \ .
\end{align}
Gauge invariant scalar perturbation quantities for the perturbations $\chi$,
$\partial \rho$, $\partial p$ of the field, the fluid energy density and the fluid
pressure respectively, can be constructed analogously:
\begin{align}
&\varphi(t, \vec x):=\bar \varphi(t)+\chi(t, \vec x)\ , & &\chi^{\rm (gi)}=\chi+R \dot{\bar \varphi} \left(B-R\dot E\right)\ , \label{eq:trafo_chi_etaZeit}\\
&\rho(t, \vec x):=\bar \rho(t) +\delta \epsilon(t, \vec x)\ , & &\delta \epsilon^{\rm (gi)}=\delta \epsilon+R \dot{\bar
\rho} \left(B-R\dot E\right)\ ,\label{eq:trafo_rho_etaZeit}\\
&p(t, \vec x):=\bar p(t) +\delta {\rm p}(t, \vec x)\ , & & \delta{\rm p}^{\rm (gi)}=\delta {\rm p}+R \dot{\bar
p} \left(B-R\dot E\right) \label{eq:trafo_p_etaZeit} \ .
\end{align}
The gauge invariant perturbation for a three vector like peculiar velocity becomes:
\begin{equation}\label{eq:trafo_u_etaZeit}
u_i(t , \vec x):=\bar u_i(t)+\delta u_i(t , \vec x)\ , \qquad\qquad \delta u_i^{\rm (gi)}=\delta
u_i+R \left(B-R\dot E\right)_{|i} \ .
\end{equation}
By using these quantities, it is possible to construct the gauge
invariant Einstein equations:
\begin{align}
&\ \Delta \Psi-3\dot R\left(\dot R\Phi +R\dot\Psi\right) +3\mathcal K \Psi=\frac{R^2}{4M^2}\delta
{T_0^0}^{\rm (gi)} \label{eq:gifg1} \\
&\left(\dot R\Phi +R \dot \Psi \right)_{,i}=\frac{R^2}{4M^2}\delta{T_i^0}^{\rm (gi)} \label{eq:gifg2} \\
&\left[\left(2 R \ddot R +{\dot R}^2\right)\Phi+R \dot R \dot \Phi+R^2\ddot\Psi +3R\dot R \dot\Psi-\mathcal
K\Psi+{1 \over 2}\Delta D \right]\delta^i_j-{1 \over
2}\gamma^{ik} D_{|kj}=-\frac{R^2}{4M^2}\delta{T_j^i}^{\rm (gi)} \label{eq:gifg3} \\
&\qquad  \qquad \qquad D:=\left(\Phi-\Psi\right) \qquad \gamma_{ij}:=\delta_{ij} \left[1+{1 \over 4}\mathcal
K \left( x^2+y^2+z^2 \right) \right]^{-2}  \ . \nonumber
\end{align}
In this equation $\mathcal K$ denotes the spacial curvature in the
Robertson-Walker metric. In the same manner a perturbed gauge
invariant Energy-momentum tensor tensor can be derived. Since
they couple only gravitationally, it is just the sum
of fluid and scalar field components:
\begin{align}\label{eq:EITgi_ges1}
&{}^{\rm}{\delta T_0^0}^{\rm (gi)}=\delta \epsilon^{\rm (gi)}-{\dot{\bar \varphi}}^{2}\Phi+{\dot{\bar
\varphi}}{\dot{\chi}^{{\rm (gi)}}}+V,_{\varphi} \chi^{\rm (gi)}\\
&{}^{\rm}{\delta T_i^0}^{\rm (gi)}=\frac{1}{R}\left( \bar \rho +\bar p\right)\delta u_i^{\rm (gi)}
+\frac{1}{R}{\dot{\bar \varphi}}{\chi^{\rm (gi)}_{,i}} \label{eq:EITgi_ges2}\\
&{}^{\rm}{\delta T_j^i}^{\rm (gi)}=\left(-\delta {\rm p}^{\rm (gi)}
+{\dot{\bar \varphi}}^{2}\Phi-{\dot{\bar
\varphi}}{\dot{\chi}^{\rm (gi)}}+V,_{\varphi} \chi^{\rm (gi)}\right)\delta^i_j \ .\label{eq:EITgi_ges3}
\end{align}
For further discussions the curvature $\mathcal K$ is set to zero
and it is possible to set $\Phi=\Psi$ because all spacial
off-diagonal elements vanish in the perturbed
Energy-momentum tensor. From (\ref{eq:gifg1}) and
(\ref{eq:gifg3}) one gets an equation that gives a relation
between $\Phi$ and $\chi^{\rm (gi)}$:
\begin{equation}\label{eq:inv_hydskafg4}
\ddot \Phi +(n+1)H\dot \Phi+\left(nH^2+2\dot H\right)\Phi+\frac{3-n}{3R^2}\Delta \Phi
=\frac{1}{4M^2}\left(\frac{n-6}{3}\dot{\bar \varphi}^2\Phi+\frac{6-n}{3}\dot{\bar \varphi}\dot \chi^{\rm (gi)}-\frac{n}{3}V,_{\varphi}
 \chi^{\rm (gi)}\right) \ .
\end{equation}
This equation does not depend on $\delta \epsilon^{\rm (gi)}$,
$\delta {\rm p}^{\rm (gi)}$ or $\delta u_i^{\rm (gi)}$. This fact
is quite remarkable, meaning that all dynamical information
already is encoded in the two variables $\Phi$ and $\chi^{\rm
(gi)}$ and their derivatives. The fluctuation in the hydrodynamical part
still can be obtained from the (0,0)-component of the
Einstein equations.

In addition to this equation one needs the
perturbation of the Klein-Gordon equation
\begin{equation}
{\varphi_{;\alpha}^{;\alpha}+ V,_{\varphi}=0} \
\end{equation}
in order to solve for the dynamics \cite{Mukhanov:1985}:
\begin{equation}\label{eq:inv_hydskafg5}
\ddot \chi^{\rm (gi)}+3H\dot \chi^{\rm (gi)}-\frac{1}{R^2}\Delta \chi^{\rm (gi)}+V,_{\varphi
\varphi}\chi^{\rm (gi)}-4\dot{\bar \varphi} \dot \Phi+2V,_{\varphi}\Phi=0 \ .
\end{equation}
\section{Discussion}
\subsection{Attractor solution I}
\begin{figure}[t!]
\centering
\psfrag{Energiedichteverh}[cc][bc][0.8]{Amplitude}
\psfrag{tau}[cc][cc]{$\tau$}
\psfrag{kinE}[cr][cr][0.7]{$\frac{\big\langle\rho^{(\varphi)}_{\rm kin, fluc}\big\rangle}
{\bar \rho^{(\varphi)}_{\rm kin}}$}
\psfrag{potE}[cr][cr][0.7]{$\frac{\big\langle\rho^{(\varphi)}_{\rm pot, fluc}\big\rangle}
{\bar \rho^{(\varphi)}_{\rm pot}}$}
\psfrag{hydE}[cr][cr][0.8]{$\frac{\left\langle\rho_{\rm hyd, fluc}\right\rangle}
{\bar \rho}$}
\psfrag{logarithmische Darstellung}[tc][tc][0.8][0]{logarithmic scaling}
\psfrag{0}[cr][cr][1][0]{$0$}
\psfrag{1}[cr][cr][1][0]{$1$}
\psfrag{2}[cr][cr][1][0]{$2$}
\psfrag{3}[cr][cr][1][0]{$3$}
\psfrag{4}[cr][cr][1][0]{$4$}
\psfrag{5}[cr][cr][1][0]{$5$}
\psfrag{6}[cr][cr][1][0]{$6$}
\psfrag{7}[cr][cr][1][0]{$7$}
\psfrag{8}[cr][cr][1][0]{$8$}
\psfrag{10}[cr][cr][1][0]{$10$}
\psfrag{0.0 10}[cr][cr][1][0]{$0$}
\psfrag{1.5 10}[cr][cr][1][0]{$1.5\cdot 10$}
\psfrag{2.5 10}[cr][cr][1][0]{$2.5\cdot 10$}
\psfrag{3.5 10}[cr][cr][1][0]{$3.5\cdot 10$}
\psfrag{4.5 10}[cr][cr][1][0]{$4.5\cdot 10$}
\psfrag{1.0 10}[cr][cr][1][0]{$1\cdot 10$}
\psfrag{2.0 10}[cr][cr][1][0]{$2\cdot 10$}
\psfrag{3.0 10}[cr][cr][1][0]{$3\cdot 10$}
\psfrag{4.0 10}[cr][cr][1][0]{$4\cdot 10$}
\psfrag{5.0 10}[cr][cr][1][0]{$5\cdot 10$}
\psfrag{6.0 10}[cr][cr][1][0]{$6\cdot 10$}
\psfrag{7.0 10}[cr][cr][1][0]{$7\cdot 10$}
\psfrag{8.0 10}[cr][cr][1][0]{$8\cdot 10$}
\psfrag{9.0 10}[cr][cr][1][0]{$9\cdot 10$}
\includegraphics[width=15cm,height=9cm]{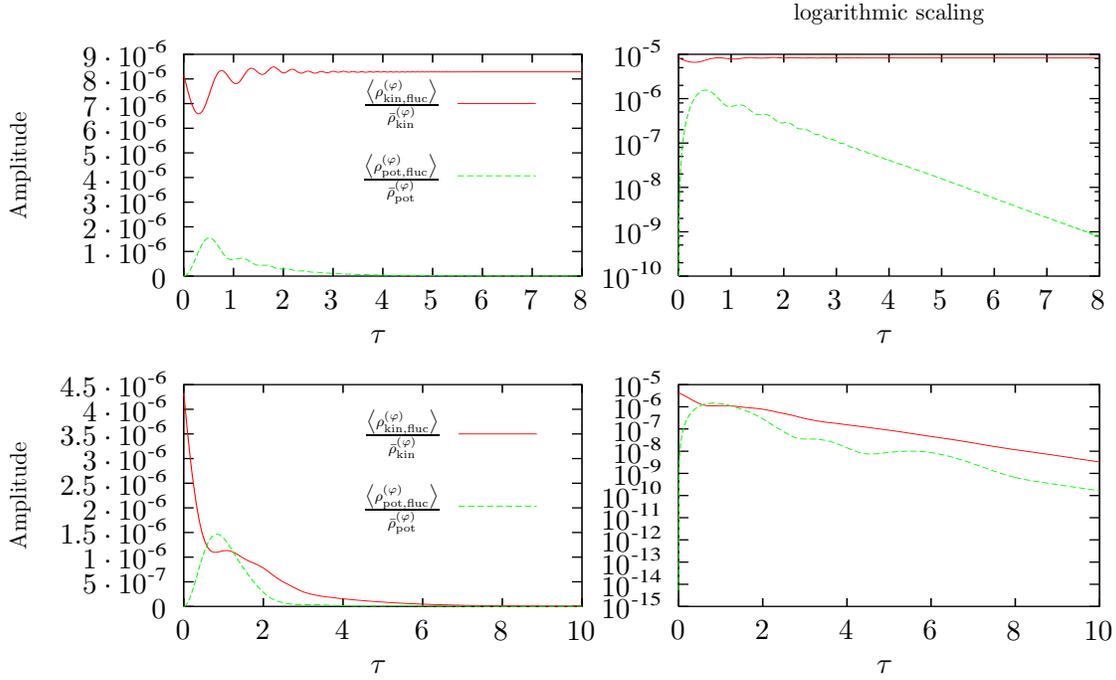}
\caption{Ratios of kinetic and potential scalar field energy
densities for attractor solution I with $a=2.5$ and $n=4$(top), $n=3$(bottom).}\label{fig:energiedichte}
\end{figure}

Equations (\ref{eq:inv_hydskafg4}) and
(\ref{eq:inv_hydskafg5}) are getting fourier-transformed (denoted
by $\tilde{  }$ ) and attractor-solution I is inserted. The
equations are rescaled into logarithmic time $\tau:=\ln
\left({t}/{t_0}\right)$ and the equations are made dimensionless
by the substitution $\tilde{\tilde \chi}:={\tilde \chi}/{M}$. The
two equations are of second order in time and it is possible to
rewrite them as a system of first order matrix ODE's:
\begin{equation}\label{eq:systemF}
\renewcommand{\arraystretch}{1.18}
\begin{pmatrix}
  {\tilde {\tilde\chi}}^{{\rm (gi)}}\\
  \tilde{\tilde\chi}^{\rm (gi)\prime}\\
  \tilde \Phi\\
  \tilde \Phi'
\end{pmatrix}'=
\renewcommand{\arraystretch}{1.18}
\begin{pmatrix}
  0 & 1 & 0 & 0 \\
  \rm f_{1} & \rm f_{2} & \rm f_{3} & \rm f_{4}\\
  0 & 0 & 0 & 1 \\
  \rm f_{5} & \rm f_{6} & \rm f_{7} & \rm f_{8}
\end{pmatrix}
\begin{pmatrix}
  {\tilde {\tilde\chi}}^{{\rm (gi)}}\\
  \tilde{\tilde\chi}^{\rm (gi)\prime}\\
  \tilde \Phi\\
  \tilde \Phi'
\end{pmatrix}
\qquad \mathrm{with}\begin{cases}
{\rm f}_{1}=\frac{2n-12}{n}-\frac{k^2}{k_{\rm R}^2}e^{(2-\frac{4}{n})\tau}& {\rm f}_{2}=\frac{n-6}{n}\\
{\rm f}_{3}=\frac{24-4n}{an}& {\rm f}_{4}=\frac{8}{a}\\
{\rm f}_{5}=\frac{6-n}{6a}& {\rm f}_{6}=\frac{6-n}{6a}\\
{\rm f}_{7}=\frac{n-6}{3a^2}+\frac{(3-n)k^2}{3k_{\rm R}^2}e^{(2-\frac{4}{n})\tau}& {\rm f}_{8}=\frac{-n-2}{n}
\end{cases}
\end{equation}
and $k_{\rm R}:=R_0 t_0^{-1}$. The prime denotes the derivative
with respect to logarithmic time. The perturbation of the
hydrodynamical part can be obtained from the (0,0)-component of the Einstein equations:
\begin{multline} \label{eq:trafo_depsilon_durch_rho2}
\frac{\tilde{\delta \epsilon}^{\rm (gi)}}{\bar \rho}=
\frac{6na-n^2a}{12a^2-6n}\tilde{\tilde\chi}^{\rm (gi)}-\frac{n^2 a}{12a^2-6n}\tilde{\tilde\chi}^{\rm (gi)\prime}
+\left( \frac{n^2-12a^2}{6a^2-3n} -\frac{n^2 a^2 k^2}{(6a^2-3n)k_{\rm R}^2}e^{(2-\frac{4}{n})\tau}\right)\tilde\Phi
-\frac{2na^2}{2a^2-n}\tilde \Phi' \ .
\end{multline}
It is common practice in cosmology to use the ensemble hypothesis
for the universe (see for instance \cite{Peacock:1999} or
\cite{Kolb:1990}) and so the $(4\times4)$-matrix needs to get
rewritten as a $(10\times10)$-matrix for the correlation
functions:
\begin{equation}\label{eq:glgsystem_allg_korr}
\frac{d}{d\tau}
\renewcommand{\arraystretch}{1.2}
{\begin{pmatrix}
  \big \langle {\tilde{\tilde \chi}}_{\vec k_1}{\tilde{\tilde \chi}}_{\vec k_2} \big \rangle\\
  \big \langle {\tilde{\tilde \chi}'}_{\vec k_1}{\tilde{\tilde \chi}'}_{\vec k_2} \big \rangle\\
  \big \langle {\tilde{\Phi}}_{\vec k_1}{\tilde{\Phi}}_{\vec k_2} \big \rangle\\
  \big \langle {\tilde{\Phi}'}_{\vec k_1}{\tilde{\Phi}'}_{\vec k_2} \big \rangle\\
  \big \langle {\tilde{\tilde \chi}}_{\vec k_1}{\tilde{\tilde \chi}'}_{\vec k_2} \big \rangle\\
  \big \langle {\tilde{\tilde \chi}}_{\vec k_1} {\tilde{\Phi}}_{\vec k_2} \big \rangle\\
  \big \langle {\tilde{\tilde \chi}}_{\vec k_1}{\tilde{\Phi}'}_{\vec k_2} \big \rangle\\
  \big \langle {\tilde{\tilde \chi}'}_{\vec k_1}{\tilde{\Phi}}_{\vec k_2} \big \rangle\\
  \big \langle {\tilde{\tilde \chi}'}_{\vec k_1}{\tilde{\Phi}'}_{\vec k_2} \big \rangle\\
  \big \langle {\tilde{\Phi}}_{\vec k_1}{\tilde{\Phi}'}_{\vec k_2} \big \rangle\\
\end{pmatrix}}=\left(
\renewcommand{\arraystretch}{1.29}
\begin{array}{cccccccccc}
  0 & 0 & 0 & 0 & 2 & 0 & 0 & 0 & 0 & 0 \\
  0 & 2\rm f_{2} & 0 & 0 & 2\rm f_{1} & 0 & 0 &  2\rm f_{3} &  2\rm f_{4} & 0 \\
  0 & 0 & 0 & 0 & 0 & 0 & 0 & 0 & 0 & 2 \\
  0 & 0 & 0 & 2\rm f_{8}& 0 & 0 &  2\rm f_{5} & 0 &  2\rm f_{6} & 2\rm f_{7}\\
  \rm f_{1} & 1 & 0 & 0 & \rm f_{2} & \rm f_{3} & \rm f_{4} & 0 & 0 & 0 \\
  0 & 0 & 0 & 0 & 0 & 0 & 1 & 1 & 0 & 0 \\
  \rm f_{5} & 0 & 0 & 0 & \rm f_{6} & \rm f_{7} & \rm f_{8} & 0 & 1 & 0 \\
  0 & 0 & \rm f_{3} & 0 & 0 & \rm f_{1} & 0 & \rm f_{2} & 1 & \rm f_{4}\\
  0 & \rm f_{6} & 0 & \rm f_{4} & \rm f_{5} & 0 & \rm f_{1} & \rm f_{7} & \rm f_{2}+\rm f_{8} & \rm
  f_{3}\\
  0 & 0 & \rm f_{7} & 1 & 0 & \rm f_{5} & 0 & \rm f_{6} & 0 & \rm f_{8}
\end{array} \right)  {\renewcommand{\arraystretch}{1.2} \begin{pmatrix}
  \big \langle {\tilde{\tilde \chi}}_{\vec k_1}{\tilde{\tilde \chi}}_{\vec k_2} \big \rangle\\
  \big \langle {\tilde{\tilde \chi}'}_{\vec k_1}{\tilde{\tilde \chi}'}_{\vec k_2} \big \rangle\\
  \big \langle {\tilde{\Phi}}_{\vec k_1}{\tilde{\Phi}}_{\vec k_2} \big \rangle\\
  \big \langle {\tilde{\Phi}'}_{\vec k_1}{\tilde{\Phi}'}_{\vec k_2} \big \rangle\\
  \big \langle {\tilde{\tilde \chi}}_{\vec k_1}{\tilde{\tilde \chi}'}_{\vec k_2} \big \rangle\\
  \big \langle {\tilde{\tilde \chi}}_{\vec k_1} {\tilde{\Phi}}_{\vec k_2} \big \rangle\\
  \big \langle {\tilde{\tilde \chi}}_{\vec k_1}{\tilde{\Phi}'}_{\vec k_2} \big \rangle\\
  \big \langle {\tilde{\tilde \chi}'}_{\vec k_1}{\tilde{\Phi}}_{\vec k_2} \big \rangle\\
  \big \langle {\tilde{\tilde \chi}'}_{\vec k_1}{\tilde{\Phi}'}_{\vec k_2} \big \rangle\\
  \big \langle {\tilde{\Phi}}_{\vec k_1}{\tilde{\Phi}'}_{\vec k_2} \big \rangle\\
\end{pmatrix}}\ .
\end{equation}
The energy density of the scalar field in second order can be
separated into an unperturbed and a perturbed part:
\begin{align}
\rho^{(\varphi)}(\dot
\varphi,\varphi) &=\frac{1}{2}\dot{\varphi}^2+\frac{1}{2}R^2\left(\partial_i\varphi
\right)^2+V(\varphi) \\
&=\underbrace{\frac{1}{2}{\dot{\bar \varphi}}^2}_{\bar \rho^{(\varphi)}_{\rm
kin}}+\underbrace{\dot{\bar \varphi}\dot \chi+\frac{1}{2}{\dot \chi}^2}_{:=\rho^{(\varphi)}_{\rm
kin, fluc}}+\underbrace{V(\bar \varphi)}_{\bar \rho^{(\varphi)}_{\rm
pot}}+\underbrace{V'(\bar \varphi){\chi}+\frac{1}{2}V''(\bar \varphi){\chi}^2}_{:=\rho^{(\varphi)}_{\rm
pot, fluc}} \nonumber \ .
\end{align}
In terms of correlation functions, this takes the form:
\begin{align}\label{eq:Energiedichte_fluc_kin}
\left\langle\rho^{(\varphi)}_{\rm kin, fluc}(\vec
x,t)\right\rangle &=\frac{1}{2}\Big\langle{\dot
\chi} (\vec x,t){\dot \chi}(\vec x,t)+\frac {1}{R^2}{\partial_i \chi(\vec x,t) \partial_i \chi(\vec x,t)}\Big\rangle\\
&=\mathop{\int}\limits_{{-\infty }}^{{+\infty }}
\frac{d^3k}{(2\pi)^3}e^{i\vec k \vec
x}\mathop{\int}\limits_{{-\infty }}^{{+\infty }}\frac{1}{2}
\delta(\vec k)\left[\big \langle \dot{\tilde \chi}\dot{\tilde
\chi} \big \rangle (q,t)-\frac{1}{R^2}q_i(k_i-q_i)\big \langle
{\tilde \chi}{\tilde \chi} \big \rangle
(q,t)\right] \frac{d^3q}{(2\pi)^3} \nonumber \\
&=\frac{4\pi}{(2\pi)^6}\mathop{\int}\limits_{\kern-8.5pt {0}}^{\kern+7pt {\infty }}\frac{1}{2} \left[\big \langle \dot{\tilde \chi}\dot{\tilde \chi} \big \rangle
(q,t)q^2+\frac{1}{R^2}\big \langle {\tilde \chi}{\tilde \chi} \big \rangle
(q,t)q^4\right]dq \nonumber \\
\mathrm{and}\qquad \qquad \nonumber\\
\left\langle\rho^{(\varphi)}_{\rm pot, fluc}(\vec
x,t)\right\rangle
&=\frac{4\pi}{(2\pi)^6}\mathop{\int}\limits_{\kern-8.5pt
{0}}^{\kern+7pt{\infty }}\frac{1}{2}V''(\bar \varphi) \big \langle
{\tilde \chi}{\tilde \chi} \big \rangle(q,t)q^2dq \ .
\end{align}
The assumption of gaussian random variables and therefore
uncorrelated fourier modes was used.
In order to do a
numerical integration of the ODE`s (\ref{eq:glgsystem_allg_korr})
one needs the initial conditions for every fourier mode. It is a
prediction of inflation that the spectrum will be scale invariant
(see for instance \cite{Kolb:1990} or \cite{Peebles:1980}),
therefore the amplitude of every correlation function $k$-mode
should be the same at the time the modes are crossing the
horizon. For attractor solution I this does happen at
\begin{equation}\label{eq:horizontzeit2_tau}
\tau_{\mathrm{H}}=\ln \left(\sqrt[\frac{2}{n}-1]{\frac{k}{k_{\rm
R}2\pi \left(1-\frac{2}{n}\right)}}\right)\ .
\end{equation}
The question how to choose the overall normalization of the
spectra remains, and indeed one does not find a theoretical
answer, because the underlying quantum theory is unknown. In this
paper the main question is, how the fluctuations will
qualitatively evolve in time. Therefore, the estimate from structure
formation that the fluctuations of the radiation part compared
to the background at the time of decoupling must have been of the
order $10^{-3}$ should be sufficient \cite{Peacock:1999}. For the sum of the correlation
functions in $k$-space this corresponds to:
\begin{equation}\label{eq:sumkorr}
\left\langle \frac{\delta \epsilon_{\rm str}(\vec x) \delta \epsilon_{\rm str}(\vec x)}{{\bar \rho}_{\rm str}^2}
\right\rangle =\frac{1}{{\bar \rho}_{\rm str}^2}\frac{4\pi}{(2\pi)^6}\int dq\ q^2\ \left\langle \tilde{\delta
\epsilon}_{\rm str}(q) \tilde{\delta \epsilon}_{\rm str}(q) \right\rangle \approx 10^{-6} \ .
\end{equation}
This is taken as a normalization for the spectra.
The fluctuations in non-relativistic matter and radiation at the
time of decoupling are related by the simple equation:
\begin{equation}
\left\langle \frac{\delta \epsilon_{\rm mat}(\vec x) \delta
\epsilon_{\rm mat}(\vec x)}{{\bar \rho}_{\rm mat}^2}
\right\rangle=\left(\frac{3}{4}\right)^2\left\langle \frac{\delta
\epsilon_{\rm str}(\vec x) \delta \epsilon_{\rm str}(\vec
x)}{{\bar \rho}_{\rm str}^2} \right\rangle \ .
\end{equation}
The result of the numerical integrations is shown in figure \ref{fig:energiedichte} for $a=2.5$.
100 different logarithmically distributed $k$-modes have been
integrated from $\tau_0=0$ to $\tau=1$ with a fourth-order
Runge-Kutta-method using a stepsize of $\Delta \tau =0.01$ and a maximum
$k$-value that corresponds to the horizon at $\tau_0$. The scale
invariance has been implemented numerically.

It can be seen in the plots that during the radiation dominated
epoch (n=4) the perturbation in the kinetic energy density
settles down to a constant value compared to the background
whereas the perturbation of the potential energy dies out
exponentially. This is an interesting result because it shows
that during the radiation dominated epoch (if attractor solution
I was valid) the perturbation of the scalar field cannot be
neglected as often argued. For other values of $a$ the
qualitative behaviour is the same, but the value that the kinetic
energy density ratio settles down to highly depends on $a$, as shown in table \ref{tab:energiedichte_a}.
\begin{table}[t!]
\centering
\scriptsize
\sloppy
\renewcommand{\arraystretch}{2}
\begin{tabular}{|l||l|l|l|l|l|l|}
  \hline
$a$ & 1.414 & 2.0 & 2.5 & 4.0 & 5.0 & 10.0 \\ \hline \rule[-4mm]{0mm}{10mm}$\frac{\left\langle\rho^{(\varphi)}_{\rm kin,
fluc}\right\rangle} {\bar \rho^{(\varphi)}_{\rm kin}}$ & $1.41\cdot 10^{-13}$& $1.76\cdot 10^{-6}$ &
$8.28\cdot 10^{-6}$ & $9.32\cdot 10^{-5}$ & $2.54\cdot 10^{-4}$ & $4.66\cdot 10^{-3}$ \\ \hline
\end{tabular}
\caption{Ratios of kinetic scalar field energy densities for attractor solution I.}
\label{tab:energiedichte_a}
\end{table}
For a value of $a=5$ as favored by nucleosynthesis one already
gets ${\langle\rho^{(\varphi)}_{\rm kin, fluc}\rangle}/ {\bar
\rho^{(\varphi)}_{\rm kin}}=2.54\cdot 10^{-4}$. The direct
comparison of the fluctuation in the energy density of the scalar
field perturbation and the radiation perturbation shows that this
ratio is nearly $a$-independent and after some time takes on a
value of ${\big\langle\delta \rho^{(\varphi)}_{\rm kin}(\vec x)
\delta \rho^{(\varphi)}_{\rm kin}(\vec
x)\big\rangle}/{\big\langle\delta \epsilon(\vec x) \delta
\epsilon (\vec x)\big\rangle} \approx 1.7$. This means that the
fluctuation in the kinetic part of the scalar field energy
density is indeed greater than in the radiation energy density.
Additionally, it is remarkable that after some time the scaling
behaviour is the same for both (like for the background energy
densities). The consequences for CMBR will be discussed later.

During the matter dominated epoch (n=3) a different picture
arises from the integrations: both kinetic and potential energy
densities of the scalar field perturbation decay exponentially
compared to the background. This means that even if the
perturbation of the scalar field has been important during the
radiation dominated epoch, in the matter dominated epoch it
became negligible. Still there will be
an interesting effect on the growing perturbation modes in the
matter density.
\subsection{Attractor solution II}
\begin{figure}[t!]
\centering
\psfrag{Energiedichteverh}[cc][bc][0.8]{Amplitude}
\psfrag{tau}[cc][cc]{$\tau$}
\psfrag{kinE}[cr][cr][0.7]{$\frac{\big\langle\rho^{(\varphi)}_{\rm kin, fluc}\big\rangle}
{\bar \rho^{(\varphi)}_{\rm kin}}$}
\psfrag{potE}[cr][cr][0.7]{$\frac{\big\langle\rho^{(\varphi)}_{\rm pot, fluc}\big\rangle}
{\bar \rho^{(\varphi)}_{\rm pot}}$}
\psfrag{hydE}[cr][cr][0.8]{$\frac{\left\langle\rho_{\rm hyd, fluc}\right\rangle}
{\bar \rho}$}
\psfrag{logarithmische Darstellung}[tc][tc][0.8][0]{logarithmic scaling}
\psfrag{0}[cr][cr][1][0]{$0$}
\psfrag{1}[cr][cr][1][0]{$1$}
\psfrag{2}[cr][cr][1][0]{$2$}
\psfrag{3}[cr][cr][1][0]{$3$}
\psfrag{4}[cr][cr][1][0]{$4$}
\psfrag{5}[cr][cr][1][0]{$5$}
\psfrag{6}[cr][cr][1][0]{$6$}
\psfrag{7}[cr][cr][1][0]{$7$}
\psfrag{8}[cr][cr][1][0]{$8$}
\psfrag{10}[cr][cr][1][0]{$10$}
\psfrag{0.0 10}[cr][cr][1][0]{$0$}
\psfrag{1.0 10}[cr][cr][1][0]{$1\cdot 10$}
\psfrag{2.0 10}[cr][cr][1][0]{$2\cdot 10$}
\psfrag{3.0 10}[cr][cr][1][0]{$3\cdot 10$}
\psfrag{4.0 10}[cr][cr][1][0]{$4\cdot 10$}
\psfrag{5.0 10}[cr][cr][1][0]{$5\cdot 10$}
\psfrag{6.0 10}[cr][cr][1][0]{$6\cdot 10$}
\psfrag{7.0 10}[cr][cr][1][0]{$7\cdot 10$}
\psfrag{8.0 10}[cr][cr][1][0]{$8\cdot 10$}
\psfrag{9. 10}[cr][cr][1][0]{$9\cdot 10$}
\includegraphics[width=15cm,height=9cm]{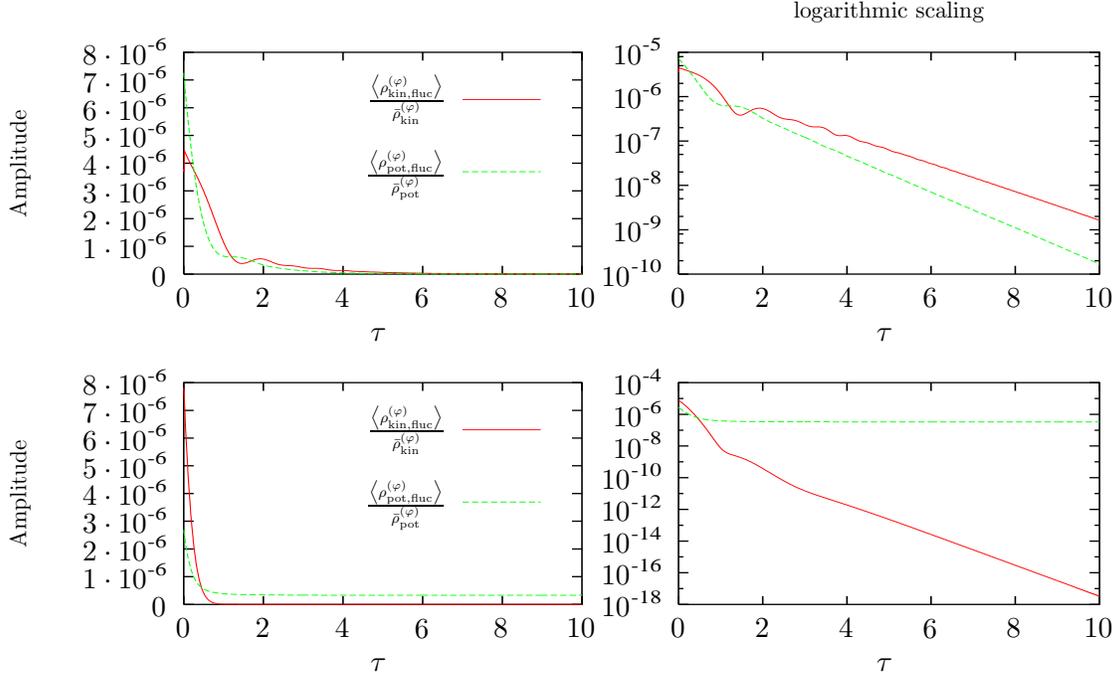}
\caption{Ratios of kinetic and potential scalar field energy
densities for attractor solution II with $a=1.2$(top) and $a=0.8$(bottom).}\label{fig:energiedichte2}
\end{figure}

The gauge invariant perturbation equations can be written in an
analogous form like (\ref{eq:systemF}) but with coefficients:
\begin{align}
\rm f_{1}&=2-\frac{6}{a^2}-\frac{k^2}{k_{\rm R}^2}e^{(2-\frac{2}{a^2})\tau}, \ &
\rm f_{2}&=1-\frac{3}{a^2}, \ &
\rm f_{3}&=\frac{12}{a^3}-\frac{4}{a}, \ &
\rm f_{4}&=\frac{8}{a}, &\\
\rm f_{7}&=\frac{n}{3 a^2}-\frac{n}{a^4}-\frac{(n-3) k^2}{3 k_{\rm R}^2}e^{(2-\frac{2}{a^2})\tau}, \ &
\rm f_{6}&=\frac{1}{a}-\frac{n}{6 a}, \ &
\rm f_{5}&=\frac{n}{2 a^3}-\frac{n}{6 a}, \ &
\rm f_{8}&=-\frac{n+1}{a^2}+1. \ & \nonumber
\end{align}
Furthermore, the equation for the time of horizon crossing is changing compared to (\ref{eq:horizontzeit2_tau}):
\begin{equation}\label{eq:horizontzeit1_tau}
\tau_{\mathrm{H}}=\ln \left(\sqrt[\frac{1}{a^2}-1]{\frac{k}{k_{\rm
R}2\pi \left(1-\frac{1}{a^2}\right)}}\right)\ .
\end{equation}
After these two alterations the equations can be integrated
analogously and the energy density of the scalar field
perturbations can be calculated. Attractor solution II becomes
valid for $a<3/3$ and there is a qualitative difference in the
behaviour for $a>1$ and $a<1$ (for $a=1$ the time dependence of
the coefficients and therefore the $k$-dependence vanishes and
this case is rather uninteresting). These two different cases are
plotted exemplary for $a=1.2$ and $a=0.8$ in figure
\ref{fig:energiedichte2}.
One sees that the ratios of the energy densities for $a > 1$
decrease exponentially, while for $a < 1$ this does happen just
for the kinetic energy density ratio. The perturbation in the
potential energy density of the scalar field settles down to a
constant value. Naively, this might explain a part of the missing
dark matter, as being stored in the local fluctuations of the
scalar field. Attractor solution II could explain today's
accelerated universe, as already mentioned. But there then must
have been a transition from attractor solution I in recent times,
meaning a long time after the beginning of matter domination.
During matter domination the ratio of the energy densities in the
scalar field dies out exponentially. Therefore, the value of the
potential energy density perturbation of the scalar field should
be very small when attractor solution II becomes valid, and
consequently the energy density stored in the perturbations is
negligible.
\subsection{Possible Consequences for CMBR and structure formation}
\begin{figure}[t!]
\centering
\psfrag{Energiedichteverh}[bc][bc][0.8][270]{$\frac{\left\langle \delta \epsilon_{\rm rad} \delta \epsilon_{\rm rad}  \right\rangle}
{\bar \rho^2}(k)$}
\psfrag{Horizont}[cl][cl][0.8]{horizon}
\psfrag{spek 1 hydro}[cr][cr][0.8]{std.-cosm. $r=-0.5$}
\psfrag{spek 1 skhydro}[cr][cr][0.8]{quintess. $r=-0.5$}
\psfrag{spek 2 hydro}[cr][cr][0.8]{std.-cosm. $r=-0.2$}
\psfrag{spek 2 skhydro}[cr][cr][0.8]{quintess. $r=-0.2$}
\psfrag{spek 3 hydro}[cr][cr][0.8]{std.-cosm. $r=0$\ \ \ \ \ }
\psfrag{spek 3 skhydro}[cr][cr][0.8]{quintess. $r=0$\ \ \ \ \ }
\psfrag{k}[cc][cc][0.8]{$k$}
\includegraphics[width=10cm,height=7cm]{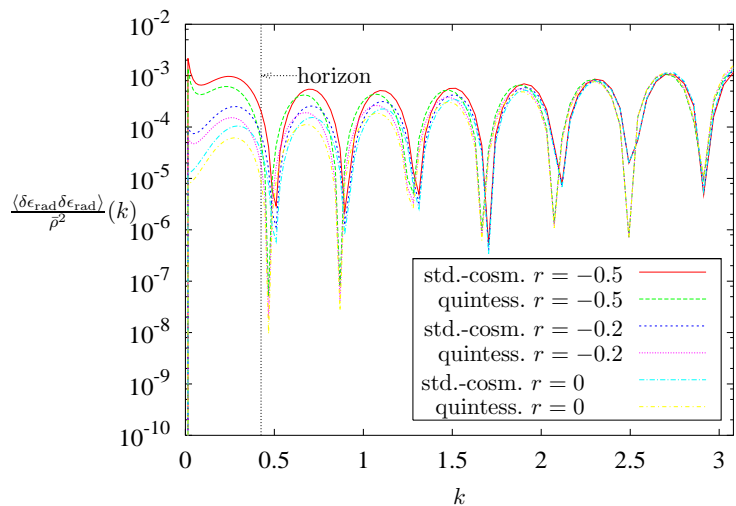}
\caption{Comparison of the spectra of standard cosmology and the
quintessential model with attractor solution I
(normalized, $n=4$, $\tau=4$, $a=2.5$)}\label{fig:sumkorrspek1}
\end{figure}

The last scattering of the CMBR occurred approximately at
decoupling. Therefore it should be generally possible to detect
the effect of the scalar field perturbation from the fluctuation
in the radiation part. It is instructive to directly compare
standard cosmology with perfect fluid plus scalar field. The main
problem is obvious: the initial normalization of the fluctuations
has been chosen to approximately fit the fluctuations of the CMBR
(\ref{eq:sumkorr}), so it would only be possible to detect a
difference between the two models in the energy density if the
qualitative behaviour of the radiation fluctuation ratio would be
different. Therefore, also the case of standard cosmology has
been examined in the ``language'' of the differential equations
(\ref{eq:systemF}) and correlation functions
(\ref{eq:glgsystem_allg_korr}). The result is that indeed one
cannot recognize a qualitative difference between this and the
case with scalar field and attractor solution I. This is rather
discouraging but there is still the hope that there might be a
difference in the $k$-spectra. To investigate this possibility
the condition of scale invariance is given up at this time and
different initial spectra are considered. Naive power spectra of
the form
\begin{equation}\label{eq:ansatz_P2}
P_{s_i s_j}(k,\tau)\propto k^{-\frac{r}{{\frac{2}{n}-1}}} e^{r\tau}
\end{equation}
are used, the free parameter is $r$. They are compared to the
respective spectra without scalar field in figure \ref{fig:sumkorrspek1}.
The difference between the two models is smaller than the
difference between different initial spectra and is getting
even smaller the earlier the modes have crossed the horizon
to the inside (corresponding to greater values of $k$). The
figure shows this for $\tau=4$, for increasing time the
differences are also getting smaller. Therefore, one
cannot distinguish the spectra of the fluctuation in the
radiation density of the two models from these naive
considerations.

One also wants to consider the influence that the existence of
the scalar field and its fluctuations have on structure
formation. The root for structure formation is that
during matter domination ($n=3$) the fluctuations of the matter density
are growing compared to the background. This is already
known from standard cosmology, and it also can be seen in the
integrations of this paper for $\varphi \equiv 0$, the growth of the correlation
functions ${\left\langle \delta \epsilon_{\rm mat} (\vec x) \delta
\epsilon_{\rm mat} (\vec x) \right\rangle}/{\bar \rho^2}$ is proportional
to $\exp(y \tau)$ with $y=4/3$ (for $\delta \epsilon_{\rm mat}/{\bar \rho}$ and
normal time this is the familiar $t^{2/3}$ behaviour of the growing modes). In the model
with scalar field and attractor solution I $(n=3)$ the growth
exponent $y$ is changed. This change is $a$-dependent as shown in table \ref{tab:fitwachstum}.
\begin{table}[t!]
\centering
\sloppy
\scriptsize
\renewcommand{\arraystretch}{3}
\begin{tabular}{|l||l|l|l|l|l|l|l|}
  \hline
$a$ & 1.414 & 2.0 & 2.5 & 4.0 & 5.0 & 10.0 & Std.-Cosmology($a\rightarrow\infty$)\\ \hline $y$ &
0.5478& 0.9996 & 1.1282 & 1.2558 & 1.2838 & 1.3204 & 1.3324\\ \hline
\end{tabular}
\caption{Fitted growth exponent $y$ for matter density in matter dominated universe($n=3$), accuracy $\sim 10^{-3}$.}
\label{tab:fitwachstum}
\end{table}
Therefore, the growth of the matter density fluctuations is
smaller than in standard cosmology for all values of $a$ and
which should lie in the region between $3/2$ and $\sim 5$ for
structure formation as discussed earlier.
\section{Conclusions and Prospects}
There are two important cases where inhomogenous fluctuations
cannot be ignored: first for attractor solution II and $a<1$
(which could play a role in late cosmology) and second in a
radiation dominated universe plus quintessence with attractor
solution I. The energy density ratio of kinetic scalar field
fluctuations compared to the background settles down to a constant
value, this numerical value highly depends on the model parameter
$a$. As a consequence, the inhomogenous fluctuation in the
radiation should have been stronger at the time of last scattering
than without scalar field. Sadly, the initial normalization of the
spectra is unknown and therefore, it is not possible to decide
between the two models from such simple considerations. The form
of the spectra also does not help very much. It might be useful to
derive the spherical multipole expansion of the fluctuations on
the one hand and to consider secondary anisotropies on the other.
But this seems to make sense only if a more realistic description
for matter and radiation is taken that pays more attention to the
different particle species, see for instance
\cite{Ma:1995ey,Ferreira:1998hj,Doran:2000jt} and a potential which is fully
compatible with observation. Furthermore, it might be interesting
to consider entropic perturbations in addition to the adiabatic
perturbations. Appropriate Ans\"atze can be found in
\cite{Mukhanov:1992me}.

The growth exponent of the growing modes for structure formation
in the matter dominated universe is changed by the coupling to
scalar field perturbations. The growth is slower than in standard
cosmology and the growth exponent highly depends on $a$. For realistic
values the growth rate changes from $\exp(4/3 \tau)$ to between
$\exp(1.28 \tau)$ for $a=5$ or $\exp(0.6 \tau)$ for $a=3/2$. This
is a strong effect and it produces the unpleasant result that the
structure formation should have taken a longer time for
quintessence with exponential potential and the tracker solution.
This seems to be a strong criteria to rule out or confirm quintessential models.
Further consequences should
be investigated for this model and also for other potentials.\bigskip

\centerline{\bf Acknowledgements} \noindent I wish to thank C.
Wetterich for useful discussions, helpful remarks and reading the
manuscript and A. Hebecker for further prospects. The main part
of this work has been done at Institut f\"ur Theoretische Physik,
Philosophenweg 16, D-69120 Heidelberg. This work is supported by DFG.
\bibliographystyle{utphys2}

\bibliography{bibtott}
\end{document}